\documentclass[aps,prb,twocolumn,superscriptaddress]{revtex4}
\usepackage{amsmath}
\usepackage{bm}
\usepackage{graphicx}
\usepackage{amsfonts}
\usepackage{amssymb}

\begin{document}

\title{Photon-assisted transport in a carbon nanotube}
\author{P. A. Orellana}
\affiliation{Departamento de F\'{\i}sica, Universidad Cat\'{o}lica del Norte, Casilla
1280, Antofagasta, Chile}
\author{M. Pacheco}
\affiliation{Departamento de F\'{\i}sica, Universidad T\'ecnica F.
Santa Mar\'{\i}a, Casilla 110-V, Valpara\'{\i}so, Chile}
\date{\today}

\begin{abstract}
We investigate the quantum transport through a single-wall carbon
nanotube connected to leads in the presence of an external radiation
field.  We analyze the conductance spectrum as a function of the
frequency and  strength of the field. We found that above a critical
value of the  field intensity, an enhancement of the conductance, or
suppressed resistance, as a function of the field strength occurs.
The conductance increases displaying oscillations which amplitude
shows a strong dependence on the field frequency. For low radiation
energies in comparison to the lead-CNT coupling energies, the
oscillations evolve toward a structure of well defined steps in the
conductance. We have shown that in this range of frequencies the
field intensity dependence of the conductance can give direct
information of single-walled carbon nanotubes energy spectra.
\end{abstract}

\maketitle

\section{Introduction}

The  electronic  transport  through carbon nanotubes (CNT's) has
received much attention in the last decade due to the  peculiar
features  of the band structure of these quasi-one dimensional
systems\cite{ijima,mildred}. Depending of their diameter and
chirality CNT's can exhibit  metallic or semiconducting behavior and
therefore be promising candidates for new carbon nanotube-based
electronic devices. Some of them have already been realized, such as
field effect transistors\cite{tans1,martel}, field emission
displays\cite{Choi} and nanosensors\cite{Cui}.

The quantum-mechanic behavior of the electronic transport in CNT's
has been experimentally confirmed by Tans et. al. \cite{tans} whom
showed that individual single-walled carbon nanotube (SWCNT) between
two contacts behaves as coherent quantum wires, and by Frank et. al.
\cite{Stefan}, they have proved the quantization of the conductance
of multiwalled carbon nanotubes. On the other hand, effects of
time-dependent potentials on transport properties of CNT's has been
studied previously for various authors \cite{frank,pan,zhao}.
Recently, Kim et. al.\cite{kang} studied experimentally the
microwave response of individual multiwall CNT finding an
enhancement of the linear conductance under the microwave radiation.

In this work, we investigate the quantum transport through SWCNT
connected to leads in the presence of an external radiation field.
Specifically, we consider a time-dependent spatially uniform
potential applied normal to the tube for modeling the effect of the
radiation field. This problem is closely related to photon-assisted
tunneling in nanostructures\cite{platero}. Basically, the external
field induces the apparition of side-bands in the spectrum and
therefore the tunneling current is drastically modified\cite{tien}.

We solve the problem using standard nonequilibrium Green's function
(NGF) techniques. The conductance is calculated by the Landauer
formula in terms of the transmission function which is obtained from
the retarded and advanced Green's function of the SWCNT in the
presence of the field, and the coupling of the nanotube to the
leads\cite{orli}. We analyze the conductance spectra as a function
of the frequency and amplitude of the external time-varying
potential. We found that above a critical value of the radiation
field intensity, an enhancement of the conductance as a function of
the  field strength occurs. The conductance increases displaying
oscillations with amplitudes strongly dependent on the field
frequency. For low photon energies in comparison to the lead-CNT
coupling energies, the oscillations evolve to a structure of well
defined steps. This effect can be explained as due to the
electric-field induced side-band resonances that increase the LDOS
at the Fermi energy opening new channels for electronic
transmission.
\section{Model}

\begin{figure}[h]
\centerline{\includegraphics[width=6cm]{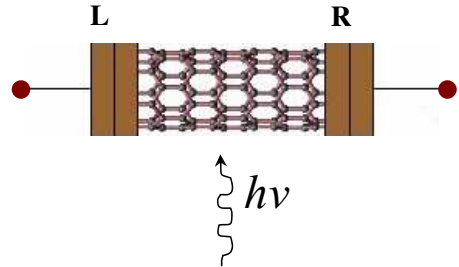}}
\caption{Schematized view of the CNT system considered.}
\label{fig1}
\end{figure}
The system under consideration is formed by a  SWCNT embedded
between two leads. The full system is modeled by the following
Hamiltonian within a noninteracting picture, that can be written as
\begin{equation}
H=H_{L}+H_{CN}+H_{LCN},
\end{equation}
with
\begin{eqnarray}
H_{L}&=&\sum_{q,\alpha}\,\varepsilon_{q_{\alpha}}(d_{q_{\alpha}}^{\dagger
}d_{q_{\alpha}}),  \notag \\
H_{CN}&=&\sum_{k}\,\varepsilon_{k}(c_{k}^{\dagger }c_{k}),  \notag \\
H_{LCN}&=&\sum_{n=1,\alpha=l,r}^{N}(V_{q_{\alpha},k}d_{q\alpha}^{%
\dagger}c_{k}+V_{q_{\alpha,k}}^{*}c_{k}^{\dagger}d_{q\alpha}),
\end{eqnarray}
\noindent where $c_{k}^{\dagger}$ is the creation operator of an
electron at the state $k$ of the carbon nanotube, and
$d_{q,\alpha}^{\dagger}$ is the
corresponding operator of an electron in the state $q$ of the right ($%
\alpha=R$) or left ($\alpha=L$) lead. Here $\varepsilon_{k}$ denotes
the energy spectrum of a metallic carbon nanotube\cite{mildred}.

 We will solve the problem using standard
non-equilibrium Green's function techniques. In this formalism the
retarded and correlated Green's function can be expressed by $
G_{k}^{r}(t,t^{\prime})=-i\theta(t-t^{\prime})\langle\{c_k(t),c_k^{\dag}(t^{%
\prime}) \} \rangle $ and $ G_{k}^{<}(t,t^{\prime})=-i\langle
c_k(t^{\prime})c_k^{\dag}(t)\rangle $. Where the Green's functions
$G^{r}$ and $G^{<}$ are obtained from the Dyson equation
$G^{r}=[(g^{r})^{-1}-\Sigma^{r}]$, and the Keldysh equation
 $G^{<}=G^{r}\Sigma^{<}G^{a}$. Where $g^{r}$ is the unperturbed Green's function of the nanotube.
  The self energies are given by,$\Sigma^{r}=-i/2(\Gamma_L+\Gamma_R)$ and
  $\Sigma^{<}=-i/2(f_L\Gamma_L+f_R\Gamma_R)$ and  $f_{\alpha }\left(
\varepsilon \right) $ is the Fermi distribution. The  line-width
functions $\Gamma _{\alpha }\left( \varepsilon \right) $, in the
wide bandwidth approximation, are taken to be independent of the
energy and energy levels. Once the Green's function $G^{r}$ and
$G^{<}$ are known, an  expression for the current can be
derived\cite{jauho}:
 \begin{widetext}
\begin{equation}
I_{\alpha }=\frac{-2e}{\hbar}\int dt^{\prime }\int
\frac{d\varepsilon}{2\pi}\Im m
\left(\sum_{k}\left\{e^{-i\varepsilon\left( t^{\prime}
-t\right)/\hbar}e^{-i/\hbar\int V\left( t^{\prime \prime }\right)
dt^{\prime \prime }}\Gamma _{\alpha }\left( \varepsilon \right)
\left[ G^{<}(t,t^{\prime })+f_{\alpha }\left( \varepsilon \right)
G^{r}(t,t^{\prime })\right] \right\} \right) ,
\end{equation}
\end{widetext}

If the CNT is perturbed by a time-dependent, spatially uniform,
potential given by: $ V(t)=V_0\cos(\omega t)$,\cite{tien}  the
zero-voltage limit linear conductance will be given
by\cite{platero}:
\begin{widetext}
\begin{equation}
G=\lim_{V\rightarrow 0}\frac{\left\langle I\right\rangle }{V}=\frac{2e^{2}}{%
\hbar }\int \frac{d\varepsilon }{2\pi }\sum_{nk}J_{n}^{2}\left(
\frac{V_{0}}{\hbar\omega }\right)(- \frac{\partial f\left( \varepsilon \right) }{%
\partial \varepsilon })\Gamma _{L}G_{k}^{r}(\varepsilon )\Gamma
_{R}G_{k}^{a}(\varepsilon ),
\end{equation}
\end{widetext}
where $\left\langle I\right\rangle$ is the  time-averaged current.
For this simple model an effective density of states can be derived,

\begin{equation}
\tilde{\rho}(\varepsilon)=\sum_{n}|J_n(\frac{V_0}{\hbar\omega})|^2\rho_{0}(\varepsilon-n\hbar\omega)
\end{equation}
where $\rho_0 (\varepsilon)$ is the bare density of states
corresponding to the CNT in the absence of the perturbing potential.
In the case of a $(n,0)$  CNT in the tight-binding approximation
this density of states can be expressed in analytic form
as\cite{luis}:
\begin{widetext}
$$
\rho_0(\varepsilon)=\frac{1}{\pi}\Im m \sum_{j=0}^{2n}\left
(\frac{-2/3(\varepsilon+i0^+)}{i\pi\sqrt{16cos^2(\frac{\pi}{n}j)
-[(\varepsilon+i0^+)^2-1-2cos^2(\frac{\pi}{n}j)]^2}}\right )
$$
\end{widetext}

 The above equation can be physically
interpreted as follow: photon absorption ($n>0$) and emission
($n<0$) can be viewed as creating an effective electron density of
states at energies $\varepsilon_n= n \hbar \omega$ with a
probability given by $|J_n(\frac{V_0}{\hbar\omega})|^2$.

\section{Results}

In what follow we will adopt the parameters: $\gamma=2.75 eV$, which
is the CNT hoping integral, and
$\Gamma_L=\Gamma_R=\Gamma=0.001\gamma$, the leads-CNT coupling
parameters. Figure 2 shows the DOS of a zigzag CNT (12,0) as a
function of the energy in units of $\gamma$, for different values of
the  radiation field strength $V_0$
 and for a particular photon energy $h\nu =0.0005\gamma$ ($\nu=332.5 GHz$). It can be seen that the
 CNT pseudo-gap is strongly modified by the presence of the sideband
 resonances appearing due to the presence of the radiation field. These
 resonances are more dense near the Fermi energy for increasing values of the field
 strength. This effect has strong influence on the system
 conductance due to the opening of new channels for electronic transmission, as we will see below.

\begin{figure}[h]
\centerline{\includegraphics[width=7cm]{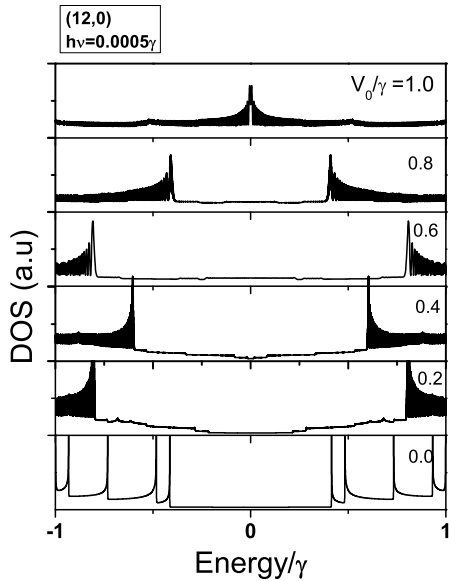}}
\caption{Density of states for a (12,0)CNT under a radiation field
of  $h\nu=0.0005\gamma$ and different values of the field strength.}
\label{fig2}
\end{figure}

\begin{figure}[h]
\centerline{\includegraphics[width=7cm]{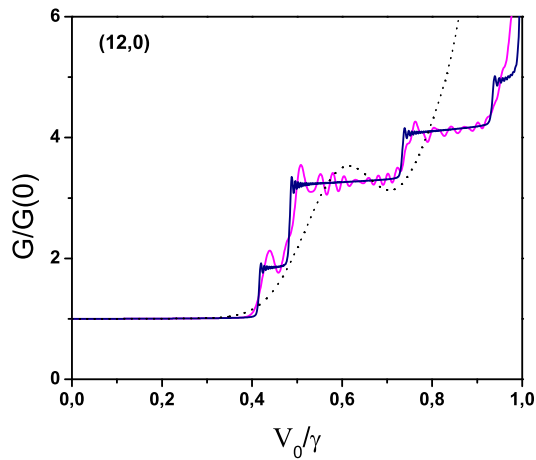}}
\caption{(color online)Normalized linear conductance of a (12,0) CNT
as a function of the oscillating field strength in unit of $\gamma$,
for $h\nu=0.0005\gamma$  (solid blue line ), $h\nu=0.005\gamma$
(solid magenta line ) and $h\nu=0.05\gamma$ (dotted line).}
\label{fig3}
\end{figure}

 In figure 3  we have displayed the  linear normalized conductance $G/G(0)$, where $G(0)$
  is the conductance for zero field,
  for the CNT (12,0) as a function of the intensity of the
 radiation field. We have plotted the corresponding conductance for  different
 photon energies
 $h\nu =0.0005\gamma$ (blue  online), $h\nu =0.005\gamma$ (magenta  online)and $h\nu =0.05\gamma$ (dotted line).
  It can be observed that in all cases the conductance increases as a function of the
radiation field strength. For low frequencies, $\nu<\Gamma/h$ the
conductance presents a very well defined structure of steps. For
increasing frequencies some oscillations appears by inducing a
complete suppression of the step features for higher frequencies.

The steps structure in the conductance for low frequencies
$\nu<\Gamma/h$, can be understood because of in that range of
frequencies the system is found in a quasi-static regime, and the
spectrum as a function of the strength of the radiation field $V_0$,
is just linearly shifted.  As the field  intensity increases the Van
Hove singularities cross the Fermi energy leading to an abrupt
increase in the conductance. In fact,  each step in the conductance
reflects the energy position of the corresponding Van Hove
singularity of the nanotube DOS. Figure 4 shows the normalized
conductance versus the strength of the radiation field for a (15,0)
CNT. There we have included  the corresponding DOS in the same
energy range.
\begin{figure}[h]
\vspace{2cm} \centerline{\includegraphics[width=8cm]{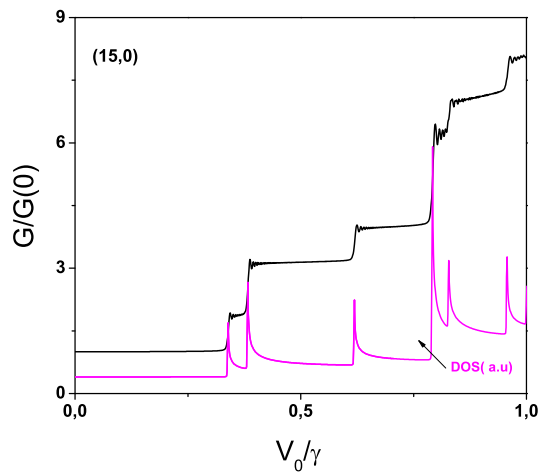}}
\caption{(color online) Normalized conductance of a (15,0) CNT as a
function of the oscillating field strength, in units of $\gamma$,
for $h\nu=0.0005\gamma$. It is also plotted the corresponding DOS of
the CNT in the same energy range. } \label{fig4}
\end{figure}
\begin{figure}[h]
\centerline{\includegraphics[width=8cm]{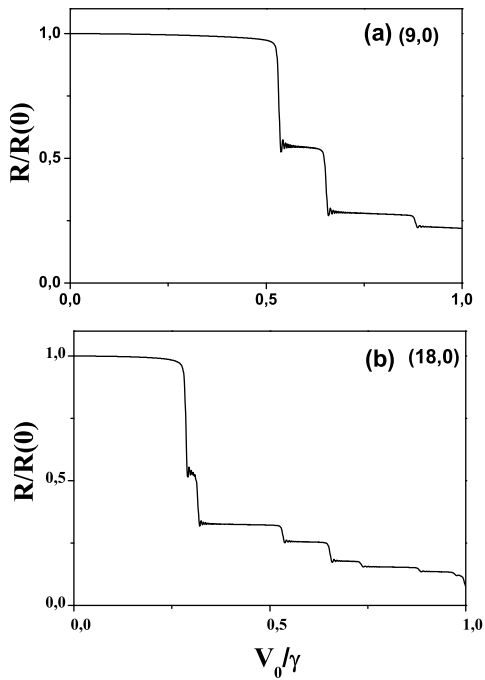}}
\caption{Normalized resistance for CNTs of different diameters as
function of the oscillating field strength, (a) (9,0) CNT and (b)
(18,0) CNT, for $h\nu=0.0005\gamma$.} \label{fig5}
\end{figure}

In figure 5 we show the behavior of  the normalized resistance
versus radiation-field strength with the nanotube radius. The figure
displays plots for (a) (9,0) and (b) (18,0) metallic nanotubes. In
both cases the step structure is clearly manifested with increasing
number of steps for larger radii. This steps structure is a reflex
of the cuasi-one dimensional density of states of the CNT.

The above results can be compared with the experimental results of
Kim at al.\cite{kang} whom studied the microwave response of an
individual  CNT. They effectively found that the resistance
decreases as a function of the radiation field power regardless of
the frequency. In their experiment the step structure is not
observed because they studied a  multiwall nanotube of about 25nm
wide. In this case the cuasi-one dimensional character of the CNT
DOS is completely suppressed.

\section{Summary}

We have investigated quantum transport through SWCNT connected to
leads in the presence of an external radiation field. We studied the
conductance spectra as a function of the frequency and of the field
strength. We found that above a critical value of the field strength
an enhancement of the conductance, or suppressed resistance, as a
function of the field intensity occurs. The conductance increases
displaying oscillations which amplitude shows a strong dependence on
the frequency of the oscillating field. For low radiation energies
in comparison to the lead-CNT coupling energies, the oscillations
evolve toward a structure of well defined steps in the conductance.
We have shown that in this range of frequencies the field intensity
dependence of the conductance can give direct information of SWCNT
energy spectra.

\section*{ACKNOWLEDGMENTS}

P.\ A.\ O.\ and M.\ P.\ would like to thank financial support from
Milenio ICM P02-054-F and FONDECYT under grants 1060952 and 1050521.

\end{document}